\documentclass[11pt,a4paper]{article}
\usepackage[backend=bibtex,maxnames=6,bibstyle=numeric,sorting=none]{biblatex}
\usepackage[utf8]{inputenc}  
\usepackage{cancel}
\usepackage[T1]{fontenc}     
\usepackage{lmodern}         
\usepackage{amsmath}         
\usepackage{amssymb}
\usepackage{amsthm}
\usepackage{bbm}             
\usepackage{bm}              
\usepackage[mathscr]{eucal}  
\usepackage{leftidx}         
\usepackage{wrapfig}
\usepackage[dvips]{graphicx}
\usepackage[usenames]{color} 
\usepackage{comment}         
\usepackage{hyperref}       

\definecolor{Black}{rgb}{0.,0.,0.}
\definecolor{Blue}{rgb}{0.,0.,1.}
\definecolor{Green}{rgb}{0.,1.,0.}
\definecolor{Cyan}{rgb}{0.,1.,1.}
\definecolor{Red}{rgb}{1.,0.,0.}
\definecolor{Magenta}{rgb}{1.,0.,1.}
\definecolor{Yellow}{rgb}{1.,1.,0.}
\definecolor{White}{rgb}{1.,1.,1.}
\definecolor{Blue4}{rgb}{0.,0.,0.5625}
\definecolor{Blue3}{rgb}{0.,0.,0.6875}
\definecolor{Blue2}{rgb}{0.,0.,0.8125}
\definecolor{LtBlue}{rgb}{0.52734375,0.8046875,1.}
\definecolor{Green4}{rgb}{0.,0.5625,0.}
\definecolor{Green3}{rgb}{0.,0.6875,0.}
\definecolor{Green2}{rgb}{0.,0.8125,0.}
\definecolor{Cyan4}{rgb}{0.,0.5625,0.5625}
\definecolor{Cyan3}{rgb}{0.,0.6875,0.6875}
\definecolor{Cyan2}{rgb}{0.,0.8125,0.8125}
\definecolor{Red4}{rgb}{0.5625,0.,0.}
\definecolor{Red3}{rgb}{0.6875,0.,0.}
\definecolor{Red2}{rgb}{0.8125,0.,0.}
\definecolor{Magenta4}{rgb}{0.5625,0.,0.5625}
\definecolor{Magenta3}{rgb}{0.6875,0.,0.6875}
\definecolor{Magenta2}{rgb}{0.8125,0.,0.8125}
\definecolor{Brown4}{rgb}{0.5,0.1875,0.}
\definecolor{Brown3}{rgb}{0.625,0.25,0.}
\definecolor{Brown2}{rgb}{0.75,0.375,0.}
\definecolor{Pink4}{rgb}{1.,0.5,0.5}
\definecolor{Pink3}{rgb}{1.,0.625,0.625}
\definecolor{Pink2}{rgb}{1.,0.75,0.75}
\definecolor{Pink}{rgb}{1.,0.875,0.875}
\definecolor{Gold}{rgb}{1.,0.83984375,0.}
\newcommand{\ist}[1]{\overset{\footnotesize(\ref{#1})}{=}}
\newcommand{\iist}[2]{\overset{^{(\ref{#1})}}{\underset{^{(\ref{#2})}}{=}}}

\newcommand{\muss}[0]{\overset{!}{=}}

\newcommand{\gleichw}[1]{\stackrel{\footnotesize(\ref{#1})}{\Leftrightarrow}}

\theoremstyle{definition}

\theoremstyle{plain}

\author{Manfried Faber,\\ Technische Universität Wien, Atominstitut,\\Stadionallee 2, 1020 Wien, Austria\\faber@kph.tuwien.ac.at}
\title{Conclusions Not Yet Drawn from the Unsolved 4/3-Problem{---}How to Get a Stable Classical Electron}
\date{}

\begin{document}


\maketitle

\begin{abstract}
It has been known for over 100 years that there is a discrepancy between Maxwell's electrodynamics and the idea of a classical electron as the ``atom'' of electricity. This incompatibility is known under the terms 4/3 problem of the classical electron and radiation reaction force and was circumvented in the currently most successful theories, the quantum field theories, by limit value considerations, by the mutual subtraction of infinities, i.e., by purely mathematical methods that eliminate obvious contradictions but are not really based on an intuitive understanding of its physical causes. The actual origin of the problems mentioned lies in the instability of the classical electron. Stabilization cannot be achieved within the framework of Maxwell's electrodynamics. This raises the question of what a minimal change to the fundamentals of electrodynamics should look like, which contains Maxwell's theory as a limiting case. A detailed analysis of the 4/3 problem points to models that fulfill these requirements.
\end{abstract}

\section{Introduction}\label{sec:Introduction}
The discussion about the origin of the problems of the classical electron is essentially about the concept of particles. The question is whether the idea of assuming atoms of electricity to describe electrodynamic phenomena, as Helmholtz had already suggested, is expedient. Stoney suggested the name ``electron'' for these ``atoms''. In formulating this description, classical electrodynamics encountered two unsolvable problems, the \mbox{4/3 problem} and the problem of radiation reaction force~\cite{Abraham1903,Lorentz1909}. In this article, we focus in particular on the cause of the 4/3 problem and examine what conclusions can be drawn from the form of the discrepancy and what kind of models can solve both problems of classical electrodynamics.

Early on in the formulation of the dynamics of electrons, an idea emerged that is still generally accepted today: a distinction is made between the dynamics of electromagnetic fields, the dynamics of electrons and the interaction between particles and fields. According to the special theory of relativity, the mass of particles is expected to increase with velocity according to the well-known Equation~(\ref{mvonv}) with the $\gamma$ factor. It is a characteristic of moving particles that are described in a ``stationary'' three-dimensional Euclidean reference system $\Sigma$, see Section~\ref{sec:Ergebnis} and Figure~\ref{mitbewegt}, that the velocity vector $u^\mu\ist{UVonbeta}\gamma(c,\vec v)$ does not have to be orthogonal to the position vectors $x^\mu:=(0,\vec x)$, i.e., the scalar product~\footnote{We use the metric $\eta^{\mu\nu}:=\mathrm{diag}(1,-1,-1,-1)$ here, i.e., $u_\mu x^\mu:=u_0x^0-\vec u\vec x$.}
\begin{equation}\label{NichtOrtho}
u_\mu x^\mu:=-\gamma\vec v\vec x\le0.
\end{equation}
does not have to vanish. The equal sign only applies to particles at rest.

For electrons, which are described by electromagnetic fields, such a particle behavior was already expected by Lorentz and Abraham~\cite{Abraham1905,Lorentz1909}. Rohrlich clearly demonstrated in~\cite{rohrlich1960,rohrlich2007} that the formalism of special relativity only guarantees that the four-momentum of a field distribution behaves Lorentz-covariantly. However, the 4/3 problem already established by Abraham~\cite{Abraham1903} shows that it was not possible to consistently represent electrons moving in $\Sigma$ by fields. The reason lies in the instability of the classical electron. To discuss the 4/3 problem, it is sufficient to consider electrons moving at constant speed. The result of such a field description of the classical electron is given in Section~\ref{sec:Ergebnis} in order to be able to discuss in Section~\ref{SecFolgerungen} which minimal changes lead to a stable model of classical electrons, so that the energy--momentum relationships also apply in reference systems that are not orthogonal to $u^\mu$, as the inequality~(\ref{NichtOrtho}) allows. In such a model, which allows a stable classical electron to be formulated, no divergences occur. Maxwell's electrodynamics would thus no longer contradict Millikan's famous experiment, which proved the quantization of the electric charge before quantum mechanics moved quantization to the center of scientific interest, as explained in Section~\ref{SecFolgerungen}.

\section{Particle and Field Description of the Classical Electron}\label{sec:Ergebnis}
Particles are lumps of matter that remain undestroyed when scattered with sufficiently low energies. Such particles can be assigned an invariant mass $m_0$ and the concepts of kinematics can be applied without contradiction. To understand what this means, it is helpful to look at the definitions and relationships of relativistic kinematics and their relationship to the non-relativistic terms, see Appendix~\ref{SecKinPunkt}.

From the assignment of the space--time coordinates $x^\mu$ to time $t$ and the position vector~$\vec x$
\begin{equation}\label{DefRaumZeit}
x:=(c_0t,\vec x)
\end{equation}
and the four-momentum $p^\mu$ to the energy $E$ and the spatial momentum $\vec p$
\begin{equation}\label{pVierer}
p:=(\frac{E}{c_0},\vec p)
\end{equation}
it follows that the mass $m$ of the particles depends on the ratio of their velocity $v$ to the speed of light $c_0$
\begin{equation}\label{mvonv}
m(\beta)\ist{pvonbeta}\gamma m_0\quad\textrm{with}\quad\gamma:=
\frac{1}{\sqrt{1-\beta^2}}\quad\textrm{and}\quad\beta:=\frac{v}{c_0}
\end{equation}
so the four-momentum results in
\begin{eqnarray}\label{vabhViererImpuls}
p\ist{pvonbeta}\gamma m_0(c_0,\vec v).
\end{eqnarray}

\begin{figure}[h]
\centering
\includegraphics[scale=0.85]{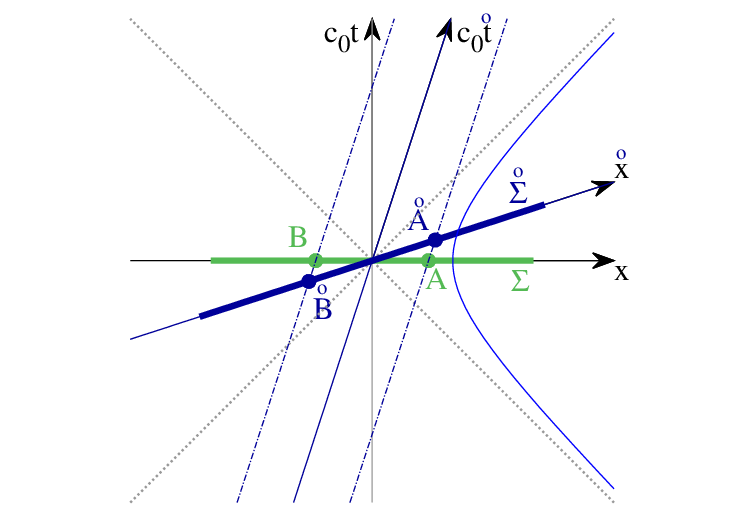}
\caption{We describe an electron in the laboratory system $\Sigma$ and in the comoving system $\stackrel{\circ}{\Sigma}$ according to the rules of special relativity, as shown in this Minkowski diagram. Note that the points $A$ and $\stackrel{\circ}{A}$ connected by dashed lines have the same spatial coordinates $\stackrel{\circ}{x}$ and only differ in time $\stackrel{\circ}{t}$. Since the electron is at rest in the comoving system, the field strengths measured in the comoving system are the same at $A$ as in $\stackrel{\circ}{A}$. To determine the field strength measured in $\Sigma$ at $A$, the coordinates $x$ and $t$ at $A$ must be taken into account and the field strength tensor must be transformed according to the transformation rules of a tensor. The relationships valid for $A$ apply accordingly for $B$.}
\label{mitbewegt}
\end{figure}

A closer look shows that  definition~(\ref{pVierer}) follows from  definition (\ref{DefRaumZeit}) if a suitable action function for a free particle is defined and the momentum is derived from it as the temporal component of the energy--momentum tensor
\begin{equation}\label{ImpulsAbgeleitetA}
p^\mu\ist{ImpulsAbgeleitet}\int_\Sigma\Theta^{\mu0}(x)\,\mathrm d^3\sigma.
\end{equation}
{The} 
 integration here takes place over that three-dimensional space-like volume $\Sigma$ in which the velocities $\vec v$ are determined, i.e., in principle over any three-dimensional space-like volume. Precisely this arbitrariness of $\Sigma$ is obviously one of the characteristics of a particle, see Figure~\ref{mitbewegt}.

In the field description,  we assume, like Abraham~\cite{Abraham1903}, that the mass of the electron is purely electromagnetic in nature, see Appendix~\ref{SecElektronMaxw}, and calculate energy and momentum according to Equation~(\ref{ImpulsAbgeleitetA}) for the field of a charge $e=-e_0$ from the symmetric energy--momentum tensor
\begin{equation}\label{symmEIT}
\Theta^{\mu\nu}\ist{symmEITEdyn}-\frac{1}{\mu_0}\eta^{\mu\kappa}F_{\kappa\lambda}F^{\nu\lambda}
+\frac{1}{4\mu_0}\eta^{\mu\nu}F_{\kappa\lambda}F^{\kappa\lambda}.
\end{equation}
{In} a reference frame $\Sigma$, in which the electron moves at a speed $\vec v=c\vec\beta$, see Figure~\ref{mitbewegt}, the field strengths transform according to the Lorentz transformation of the field strength tensor $F^{\mu\nu}$ to
\begin{eqnarray}\begin{aligned}\label{EBbew}
&\vec E(\vec\beta)\ist{EBbewegt}\vec E_\parallel(0)+\gamma\vec E_\perp(0)\quad
\textrm{with}\quad\vec E_\parallel:=\frac{\vec\beta(\vec\beta\vec E)}{\beta^2},
\quad\vec E_\perp:=\vec E-\vec E_\parallel,\\
&c\vec B(\vec\beta)\ist{EBbewegt}\gamma\vec\beta\times\vec E(0).
\end{aligned}\end{eqnarray}
{For} a charge at rest, each of the three electric field components contributes to the energy density~(\ref{EneDicht}) with one third of the rest energy density $\mathcal E_0(\vec r)$. This leads to an energy of the moving charge $e$
\begin{equation}\label{eneBewegt}
E_e(\beta)\ist{eneBew0}E_e(0)\,\frac{4\gamma^2-1}{3\gamma},
\end{equation}
that does not have the form expected for particles~(\ref{vabhViererImpuls}), $E_e(\beta)=\gamma E_e(0)$. The momentum
\begin{equation}\label{ImpBewegt}
\vec P_e(\vec\beta)\ist{ImpBew0}\frac{\varepsilon_0}{c_0}
\int_{\stackrel{\circ}{\Sigma}}\mathrm d^3\stackrel{\circ}{\sigma}
\left[-\vec E_\perp(0)\left(\vec E_\parallel(0)\vec\beta\right)
+\gamma\vec\beta\vec E_\perp^2(0)\right]
\ist{ImpBew1}\vec\beta\gamma\frac{4}{3}\frac{E_e(0)}{c_0}
\end{equation}
shows momentum densities normal to the velocity
\begin{equation}\label{spannungen}
-\frac{\varepsilon_0}{c_0}
\vec E_\perp(0)\left(\vec E_\parallel(0)\vec\beta\right)
\end{equation}
and thus internal stresses in the classical electron, which cancel each other out and therefore do not contribute to the total momentum. The factor 4/3 in $\vec P_e(\vec\beta)$ means a discrepancy between the gravitational mass $E_e(0)/c_0^2$ and the inertial mass $\frac{4}{3}E_e(0)/c_0^2$ of the classical electron, which can be read from expression~(\ref{ImpBewegt}). This is in obvious contrast to the particle description of a classical electron according to Equation~(\ref{vabhViererImpuls}). The cause of this contradiction lies in the instability of the classical electron described by the electromagnetic fields of Maxwell's electrodynamics, as has been known for over 100 years~\cite{poincare1905} and is explained in detail in Section~\ref{SecFolgerungen}. Furthermore, conclusions are drawn in this section as to how a model of a stable classical electron can be formulated.

The fact that the discrepancy discussed in this paper, which is over 100 years old, is not a violation of the rules of special relativity is confirmed by the calculation of the four-momentum of the moving electron, if the coordinates $(c_0t,\vec x)$ are used for the calculation, but integrated over the space-like volume $\stackrel{\circ}{\Sigma}$, which are simultaneous for the electron at rest. This integration leads to 
\begin{equation}\label{ViererSigmaErg}
P^\mu(\vec\beta)\ist{ViererSigma0}\frac{E_e(0)}{c_0}(\gamma,\vec\beta\gamma),
\end{equation}
and relates to the four-momentum of the electron at rest as expected. That the rules of special relativity also apply to the unstable classical electron was emphasized by Rohrlich with his calculations in Ref.~\cite{rohrlich2007}.

\section{Conclusions from the Problems}\label{SecFolgerungen}
An interpretation of the frustrating result~(\ref{eneBewegt}) for the energy of the moving electron is facilitated by a comparison with the Sine--Gordon model, a Lorentz-invariant, topologically interesting, field model in one space and one time dimension, which is illustrated very clearly in Ref.~\cite{remoissenet:1999wa}. According to Equation~(6.23) of Ref.~\cite{remoissenet:1999wa} the rest energy $E_\mathrm{SG}(0)$ of a Sine--Gordon soliton increases for a moving soliton {to}
\begin{equation}\label{SGBew}
E_\mathrm{SG}(\beta)=E_\mathrm{SG}(0)\frac{\gamma^2+
\overbrace{1+\gamma^2\beta^2}^{\gamma^2}}{2\gamma}
=\gamma E_\mathrm{SG}(0).
\end{equation}
{The} stress energy $\propto\gamma$, the potential energy $\propto1/\gamma$ and the kinetic energy $\propto\gamma\beta^2$ are listed here in sequence. This result, $E_\mathrm{SG}(\beta)=\gamma E_\mathrm{SG}(0)$, for the energy of the Sine--Gordon soliton fulfills the expectations of a particle that is subject to the laws of relativistic kinematics. It is also stable because the stress term broadening a particle~$\equiv$~Sine--Gordon soliton and the compressing potential energy keep each other in equilibrium. In the soliton at rest, these two energy contributions must be equal in order for stability to occur according to the Hobart--Derrick theorem~\cite{Hobart:1963rh,Derrick:1964gh}. This results from the one-dimensional integration over the real axis and the number of derivatives, as the stress term contains two derivatives and the potential term contains no derivative. The stress energy is therefore proportional and the potential energy indirectly proportional to the diameter of the soliton.

When comparing the energy expressions~(\ref{eneBewegt}) and (\ref{SGBew}), it is noticeable that adding an energy contribution of one third of the rest energy and with a $1/\gamma$ behavior, i.e., of $\frac{E_e(0)}{3\gamma}$, to the energy $E_e(\beta)$ in Equation~(\ref{eneBewegt}) leads to the behavior expected for a stabilized classical~electron
\begin{equation}\label{korrEne}
E_e(\beta)\to E_\mathrm{stab}(\beta):=E_e(\beta)+\frac{E_e(0)}{3\gamma}
\ist{eneBewegt}\frac{4}{3}\gamma E_e(0).
\end{equation}
{The} addition of this contribution thus leads to the energy value required by the momentum calculation~(\ref{ImpBewegt}) of the moving classical electron. The added energy $\frac{E_e(0)}{3\gamma}$ is obviously the energy contribution required for stabilization. After taking it into account, the energy $E_e(0)$ of the electric field is only 3/4th of the rest energy of a stable classical electron. The size of the added contribution, one third of the electromagnetic field energy for a particle at rest, shows that this contribution must not have any Lorentz indices, i.e., it must be a potential energy. This is because only such a contribution
\begin{equation}\label{potEne}
E_\mathrm{pot}(0):=\int\mathrm d^3r\,\mathcal E_\mathrm{pot}(\vec r)
\end{equation}
to the energy of a particle at rest scales under the substitution $r\to\lambda r$ as one third of the electric field energy of a particle at rest, $E_\mathrm{pot}(0)\ist{skalEnergien}E_e(0)/3$,
\begin{equation}\begin{aligned}\label{skalEnergien}
\frac{\mathrm d}{\mathrm d\lambda}&\left[\int\mathrm d^3(\lambda r)\,
\mathcal E_0(\lambda r)\right]_{\lambda=1}
+\frac{\mathrm d}{\mathrm d\lambda}\left[\int\mathrm d^3(\lambda r)\,
\mathcal E_\mathrm{pot}(\lambda r)\right]_{\lambda=1}=\\
&=-E_e(0)+3E_\mathrm{pot}(0)=0,
\end{aligned}\end{equation}
if the total energy $E_\mathrm{stab}(\beta)=E_e(\beta)+E_\mathrm{pot}(\beta)$ has a minimum at $\lambda=1$. Such an energy contribution $E_\mathrm{pot}(\beta)$ also contributes nothing to the momentum $\vec P_e(\vec\beta)$ of a moving electron.

We conclude from this comparison with the Sine--Gordon model that the solution of the 4/3 problem requires a formulation of the degrees of freedom of an electron that allows the field values that occur in the center of the electron to have a high potential energy density and thus prevent an unlimited increase in the size of the central region. This is only possible with a formulation of electrodynamics in which the vector fields $A_\mu$ are not the fundamental fields, but with vector fields based on a scalar field (Higgs field) $Q(x)$, with which a suitable potential energy density $\mathcal E_\mathrm{pot}(0)$ can be formulated, which disappears sufficiently quickly at infinity. The trick is to find a formulation in which the dynamics of the scalar field is formulated by the vector field $A_\mu$ in the usual form with four Lorentz~indices.

We can draw another important conclusion from the internal stresses that we have determined in connection with the calculation of the momentum of a moving classical electron in Equation~(\ref{ImpBewegt}). These stresses are formulated in the first term in the square brackets in Equation~(\ref{ImpBewegt}). Although they cancel each other out in the calculation of the total momentum, they reflect the instability of the classical electron. These stresses only disappear if the field strength components $\vec E_\parallel(0)$ and $\vec E_\perp(0)$ are everywhere orthogonal to each other in some additional, internal space. This can be achieved in non-abelian formulations, when the field strength tensors take values in the Lie algebra of some nonabelian group, as they occur in quantum chromodynamics or in the su(2) algebra, in which such field components can belong to orthogonal directions in the algebra.

The 4/3 problem shows that within the Maxwell-Lorentz theory it is impossible to describe electrons as particles localized in a small volume with a quantized charge whose field extends to infinity or to oppositely charged particles. It therefore follows that a modification of the theory is absolutely necessary. This modification should lead to a simple model, but what does simple mean? A simple description is characterized by its intuitiveness, i.e., its direct comprehensibility resulting from an idea or a comparison. The fewer defining fields a description contains, the fewer relations between these fields need to be chosen. As far as electrodynamics and the description of electrons are concerned, there are four scales invented by humans: time, length, mass and charge. Since 2019 they have been formulated in the SI by the numerical values of $c_0,e_0,\hbar\;\textrm{and}\;\Delta\nu$ of Cs. The conversion between the four associated natural scales of a model and the SI scales can only be obtained by comparison with the experiment. These four scale parameters are not free parameters. A simple model should require as few adjustable parameters as possible. As Einstein's description of gravitation shows, a simple model does not necessarily lead to simple calculations. Due to possible non-linearity, even a simple model may require complicated calculations.

As a result of the above analyses, I would like to set the following requirements for a simple modified classic model of the electron and electrodynamics:
\begin{enumerate}
\item There is no division of the field degrees of freedom between degrees of freedom for electrons and degrees of freedom for electromagnetic fields, no division of the Lagrangian function into a dynamics of free fields, a dynamics of free particles and an interaction term between these free fields, as exemplified by the Sine--Gordon model. In experiments, electrons are inseparable from their fields. There are no electrons without fields. This is also how electrons should be described.

\item Electrons should be purely electromagnetic in nature, so their dynamics should be describable by a maximum of three field degrees of freedom as in Maxwell's electrodynamics. Of the four $A_\mu$ fields, one of the degrees of freedom is only a gauge degree of freedom and is therefore described as unphysical. The Maxwell-Diracian description uses four fields for the photon field $A_\mu$ and 8-1 degrees of freedom for the complex, normalized four-component Dirac spinors.

\item As was concluded in connection with Equation~(\ref{korrEne}), the Lagrangian density sought should contain, in addition to the dynamic term with four derivatives that seeks to smear electrons, a potential term without derivatives that holds the electron together. The field itself, which describes the electrons, is uncharged. The problem of the instability of the classical electron, which has remained unsolved for 100 years, cannot be attributed to the repulsion of charged regions inside the electron, as is often assumed~\cite{Schwinger1983}. The topological structure of the field of charges should lead to attraction or repulsion and quantization of the charges due to the terms in the Lagrangian.

\item The dynamic term with four derivatives should asymptotically transform the field into the structure of the field of a point charge $e$, i.e., into a field that can be described in Abelian terms. The potential term should only have a short-range effect and modify the Coulomb law at small distances, i.e., cause the charge to run due to the geometry.

\item The description of electrons following from the Lagrangian should bosonize Maxwell-Dirac's formulation of electrodynamics.
\end{enumerate}

In addition to the stabilization of electrons, a potential term in the Lagrangian would allow further interesting conclusions to be drawn. It already classically leads to a non-vanishing trace of the energy--momentum tensor~\cite{wabnig25}. The ``trace anomaly'' of QED would thus find an explanation. Anomalies are symmetries of the Lagrangian, i.e., of the classical description, which are lost in quantum theory. Scale transformations lead to the vanishing of this trace in the classical description of electrodynamics for massless electrons. In QED, however, this trace remains non-zero even for vanishing fermion mass, as explained in more detail in section 19.5 of Peskin-Schröder~\cite{peskin} with reference to the original work~\cite{Callan1970}. The trace anomaly can therefore be seen as an indication of the absence of the potential term in classical electrodynamics.

QED suffers from the serious problem that it cannot explain the vacuum energy density of the universe. From the measurements of the Planck Collaboration~\cite{planck:2018},  its value is 0.69\,$\rho_\mathrm{crit}$, whereby the critical energy density is $4.9\,\frac{\textrm{GeV}}{\textrm m^3}$. The many fluctuations of the quantum fields could in principle contribute to the vacuum energy density. However, if they are taken into account, they result in a value that is dozens of powers of ten too high. There is no consensus on the value, but factors of up to 120 powers of ten are mentioned. The usual alternative in quantum field theory of subtracting vacuum expectation values leads to an error at least as large, namely the disappearance of the value. This is known as the “cosmological constant problem”. With a potential term in the Lagrangian of nonlinear electrodynamics, each particle would contribute directly to the vacuum energy density and Einstein's cosmological constant should result as the spatial average of these contributions.

There are many papers on the 4/3 problem~\cite{poincare1905,Dirac1938,Schoenberg1946,Kwal1949,Caldirola1956,rohrlich1960,Coleman1961,Dirac1962,Nodvik1964,Caldirola1978,Barut1981,Boyer1982,Ferraris1982,Bialynicki-Birula1983,Coleman1983,Schwinger1983,Visser1989,Frenkel1996,Williamson1997,Kiessling1999,Jimenez1999,rohrlich2007,Polonyi2017,Consa2018,Chubykalo2019,Landvogt2023,Klingman2024,Bini2024}, as well as other studies cited in these works. Some of them also deal with the second problem of classical electrodynamics, the radiation reaction problem. For a formulation of electrodynamics with only one non-Abelian field, following the above conditions, this second problem is irrelevant. A single fundamental non-Abelian field with a suitable Lagrangian function as proposed above cannot have a reaction on itself, but can only follow its dynamics as formulated in the Lorentz invariant Lagrangian. As far as the Lagrangian is a Lorentz scalar, no contradiction with special relativity can occur due to the consistency of the theory. Each scientist will form their own image and decide for themselves which image is most likely to be developed further. The above considerations show what minimal changes sould be made to Maxwell's electrodynamics in order to eliminate the inconsistencies, which are more than a hundred years old. Maxwell's electrodynamics should then turn out to be a clever linear approximation to this nonlinear theory. It should be interesting to investigate models with such properties~\cite{Faber:1999ia,Faber:2022zwv}.

We can then ask ourselves how fundamental theoretical physics would have developed if a stable field configuration for an electron had been found soon after the investigations of Abraham and Lorentz, for a model of a classical electron whose far field corresponds to the Coulomb field of classical electrodynamics for a point-shaped electron, as suggested by Equation~(\ref{Ereg}). There would then have been no singularity that would have had to be regularized. I think it is quite unlikely that the path proposed by Kramers at the Shelter Island Conference in 1947 would have been taken. As Veltman put it in a lecture~\cite{Veltman2015}, Kramers said: ``Well, let's do it like follows: Let's not try to understand everything. Let's just say the experimental mass, that's something we can measure and God knows what goes on in the electron at small distances and the like. Why don't we just skip that part of the problem?'' I also don't assume that the great physicists of the past and present would not have come up with anything to explain the experiments in such a model.

\appendix
\numberwithin{equation}{section}

\section{Kinematics of Point Particles \label{SecKinPunkt}}
A relativistic description of the motion of point particles requires that from the knowledge of the kinematic quantities in one inertial system, their values can be calculated in other inertial systems. It is therefore useful to formulate kinematic quantities as scalars, vectors and tensors. This allows for a simple transformation of these quantities between the reference systems and an easy check of whether the principle of relativity is fulfilled, i.e., whether the same physical laws apply in all reference systems.

The transformation
\begin{eqnarray}\label{ViererTrafo}
x^{\mu\prime}:={\Lambda^{\mu\prime}}_\nu x^\nu\quad
\Leftrightarrow\quad x^\prime:=\Lambda x,
\end{eqnarray}
of the coordinates $x^\mu\ist{DefRaumZeit}(ct,\vec x)$ defines a Lorentz transformation from the laboratory system $\Sigma$ to the moving system $\Sigma^\prime$. The condition
\begin{eqnarray}\label{DefLambdaTrafo}
\Lambda^T\eta\Lambda\muss\eta\quad\textrm{with}\quad
\eta:=\mathrm{diag}(1,-1,-1,-1)
\end{eqnarray}
on the Lorentz transformations $\Lambda$ guarantees the invariance
\begin{eqnarray}
x^\mu x_\mu\iist{ViererTrafo}{DefLambdaTrafo}x^{\mu\prime}x_{\mu\prime}\quad\textrm{with}\quad x_\mu=\eta_{\mu\nu}x^\nu
\end{eqnarray}
and the invariance of the proper time $\tau$ by the differential
\begin{eqnarray}\label{tauDiff}
\mathrm d\tau:=
\frac{1}{c}\sqrt{\mathrm dx_\mu\mathrm dx^\mu}\ist{DefRaumZeit}
\mathrm dt\sqrt{1-\Big(\frac{\mathrm d\vec x}{\mathrm dt}\Big)^2}.
\end{eqnarray}
The gradients transform ``contragrediently'' with the matrix $\bar\Lambda$
\begin{eqnarray}
\partial_{\mu\prime}\ist{ViererTrafo}\bar\Lambda_{\mu\prime}^{\ \ \nu}\partial_\nu
\quad\Leftrightarrow\quad\partial^\prime=\bar\Lambda\partial
\quad\textrm{with}\quad\bar\Lambda:=\eta\Lambda\eta
\ist{DefLambdaTrafo}{\Lambda^T}^{-1}.
\end{eqnarray}

Using the non-relativistic definitions of velocity and acceleration
\begin{eqnarray}\label{DreiGroess}
\vec v:=\frac{\mathrm d\vec x}{\mathrm dt},\quad
\vec a:=\frac{\mathrm d\vec v}{\mathrm dt},
\end{eqnarray}
and the abbreviations
\begin{eqnarray}\label{betagammatau}
\vec\beta:=\frac{\vec v}{c_0},\quad
\gamma:=\frac{1}{\sqrt{1-\beta^2}},\quad
\mathrm dt\iist{tauDiff}{DreiGroess}\gamma\,\mathrm d\tau,
\end{eqnarray}
we obtain the four-velocity 
\begin{eqnarray}\label{UVonbeta}
u:=(u^0,\vec u):=\frac{\mathrm dx}{\mathrm d\tau}=
\frac{\mathrm dx}{\mathrm dt}\frac{\mathrm dt}{\mathrm d\tau}
\iist{DefRaumZeit}{betagammatau}\gamma(c_0,\vec v)
\ist{betagammatau}c_0\gamma(1,\vec\beta),
\end{eqnarray}
and the four-acceleration, \footnote{
\begin{eqnarray}\label{gammaAbl}
\frac{\mathrm d\gamma}{\mathrm dv_i}\ist{betagammatau}
\frac{1}{c_0}\frac{\mathrm d\gamma}{\mathrm d\beta_i},\quad
\frac{\mathrm d\gamma}{\mathrm d\beta_i}
\ist{betagammatau}\frac{\beta_i}{(1-\beta^2)^{3/2}}=\beta_i\gamma^3,
\end{eqnarray}}
\begin{eqnarray}\begin{aligned}\label{bVona}
b&=(b^0,\vec b):=\frac{\mathrm du}{\mathrm d\tau}\ist{betagammatau}
\frac{\mathrm du}{\mathrm dv_i}\frac{\mathrm dv_i}{\mathrm dt}\gamma
\iist{UVonbeta}{DreiGroess}\frac{\mathrm d}{\mathrm dv_i}
\left[\gamma(c_0,\vec v)\right]a_i\gamma=\\
&\ist{gammaAbl}\left(\gamma^4\vec a
\vec\beta,\gamma^2\vec a+\gamma^4(\vec a\vec\beta)\vec\beta\right),
\end{aligned}\end{eqnarray}
which for rectilinear motion is simplified to the space-like four-vector
\begin{eqnarray}\label{bgerade}b\iist{bVona}{betagammatau}
\gamma^4a\left(\beta,\vec1\right)
\end{eqnarray}

Including the invariant rest mass $m_0$, we define the four-vectors for the momentum $p^\mu$ and the force vector $K^\mu$
\begin{eqnarray}\label{pUndK}
p:=(\frac{E}{c_0},\vec p):=m_0u,\quad
K:=(K^0,\vec K):=\frac{\mathrm dp}{\mathrm d\tau}\ist{bVona}m_0b.
\end{eqnarray}
The following applies to the four-momentum
\begin{eqnarray}\label{pvonbeta}
&&p\ist{pUndK}m_0u\ist{UVonbeta}\gamma m_0(c_0,\vec v)=:m(\beta)(c_0,\vec v)
\end{eqnarray}
and the four-force
\begin{eqnarray}\label{defK}
K\ist{pUndK}m_0b\ist{bVona}m_0\left(\gamma^4\vec a
\vec\beta,\gamma^2\vec a+\gamma^4(\vec a\vec\beta)\vec\beta\right).
\end{eqnarray}

It is noteworthy that for the derivation of the relations of the relativistic to the non-relativistic quantities of particle kinematics, in addition to the four-vector (\ref{DefRaumZeit}) of the coordinates $x^\mu$, we have also identified the four-momentum $p^\mu$ directly with the non-relativistic quantities, energy $E$ and momentum $\vec p$, see Equations~(\ref{pVierer}) and (\ref{pUndK}). We will elaborate on this after Equation~(\ref{minimalerWeg}).

From expression~(\ref{pvonbeta}) for the four-momentum $p$, it follows that the mass contributing to energy and momentum increases proportionally to $\gamma$,
\begin{eqnarray}\label{pundgamma}
E\iist{pUndK}{pvonbeta}\gamma m_0c_0^2,\quad
\vec p\iist{pUndK}{pvonbeta}\gamma m_0\vec v.
\end{eqnarray}

Since we retain the definitions for the power $P$ and the force $\vec F$
\begin{eqnarray}\label{FundP}
P:=\frac{\mathrm dE}{\mathrm dt},\quad
\vec F:=\frac{\mathrm d\vec p}{\mathrm dt}
\end{eqnarray}
known from non-relativistic mechanics, it follows that 
\begin{eqnarray}\label{3und4}
K\ist{pUndK}m_0b\ist{bVona}m_0\left(\gamma^4\vec a
\vec\beta,\gamma^2\vec a+\gamma^4(\vec a\vec\beta)\vec\beta\right)
\iist{pUndK}{betagammatau}\gamma
\left(\frac{1}{c_0}P,\vec F\right),
\end{eqnarray}
i.e.,
\begin{eqnarray}\label{Funda}
\vec F\ist{3und4}m_0[\gamma\vec a+\gamma^3(\vec a\vec\beta)\vec\beta]
\quad\textrm{und}\quad P=\vec v\vec F.
\end{eqnarray}
{For} rectilinear motion, it follows that the inertial mass that must be accelerated increases with $\gamma^3$
\begin{eqnarray}\label{m0mita}
\vec F\ist{Funda}\gamma^3m_0\vec a.
\end{eqnarray}

We will now calculate invariants and draw conclusions from them. Vectors $x^\mu$ are denoted by
\begin{eqnarray}
x^\mu x_\mu=\begin{cases}\rho^2>0\;&\textrm{timelike vectors},\\
0\;&\textrm{lightlike vectors},\\
-\rho^2<0\;&\textrm{spacelike vectors}.
\end{cases}
\end{eqnarray}
$x^\mu x_\mu$ is a Lorentz invariant but not an invariant of motion. In contrast, however,
\begin{eqnarray}
&&u_\mu u^\mu\ist{UVonbeta}c_0^2\gamma^2(1-\vec\beta^2)
\ist{betagammatau}c_0^2\label{uuInv}\\
&&p_\mu p^\mu\iist{pUndK}{uuInv}m_0^2c_0^2\label{ppInv}
\end{eqnarray}
are also invariants of motion. From the vanishing of the differential of
these invariants follows the relativistic energy conservation law~\footnote{
\begin{eqnarray}\label{ortVA}
u^0\mathrm du^0\ist{uuInv}\vec u\,\mathrm d\vec u\quad\gleichw{UVonbeta}\quad
c_0^2\mathrm d\gamma=\vec v\,\mathrm d(\gamma\vec v)\ist{DreiGroess}
\frac{\mathrm d\vec x}{\mathrm dt}\,\mathrm d(\gamma\vec v)
=\mathrm d\vec x\,\frac{\mathrm d(\gamma\vec v)}{\mathrm dt}
\end{eqnarray}}
\begin{eqnarray}\label{EnErh0}
\mathrm dE\ist{pundgamma}\mathrm d\gamma m_0c_0^2\ist{ortVA}
\mathrm d\vec x\,\frac{\mathrm d(\gamma m_0\vec v)}{\mathrm dt}\ist{pvonbeta}
\mathrm d\vec x\,\frac{\mathrm d\vec p}{\mathrm dt}\ist{FundP}
\vec F\mathrm d\vec x,
\end{eqnarray}
which expresses that mechanical work $\int\vec F\mathrm d\vec x$ contributes to the energy. If Equation~(\ref{EnErh0}) is divided by $\mathrm dt$, it turns out that energy conservation was already included in Equation~(\ref{Funda})
\begin{eqnarray}\label{MechLeist}
P\ist{FundP}\frac{\mathrm dE}{\mathrm dt}\iist{EnErh0}{DreiGroess}\vec v\vec F.
\end{eqnarray}

The two Equations (\ref{pUndK}) and (\ref{ppInv})  show that the energy of a
moving mass, in addition to the kinetic energy and its relativistic
corrections, also contains an additional contribution, the rest energy $m_0c_0^2$
\begin{eqnarray}\begin{aligned}\label{relEnergieBez}
&(p^0)^2-\vec p^{\,2}\ist{ppInv}m_0^2c_0^2\quad\Leftrightarrow\quad
E^2\ist{pUndK}m_0^2c_0^4+\vec p^{\,2}c_0^2,\\
&E\ist{pundgamma}\gamma m_0c_0^2\ist{betagammatau}\frac{1}{\sqrt{1-\beta^2}}m_0c_0^2
\approx m_0c_0^2+\frac{m_0\vec v^{\,2}}{2}+\mathcal O(\beta^4).
\end{aligned}\end{eqnarray}

Interestingly, by transferring the law of conservation of energy from classical mechanics to relativity theory, it has emerged naturally that for the two canonically conjugated quantities $x$ and $p$, the spatial components coincide with the three-quantities. However, for their derivatives $u$, $b$ and $K$, we had to introduce separate letters to prevent confusion between their spatial components and the non-relativistic quantities $\vec v$, $\vec a$ and $\vec F$.

We will now show that it is sufficient to define the relationship~(\ref{DefRaumZeit}) between three- and four-quantities. The relationship~(\ref{pVierer}) between the momenta follows from a suitably chosen Lagrange density and the energy--momentum tensor derived from it, see Equation~(\ref{ImpulsAbgeleitet}).

Free particles move on geodesics, on extremal paths between events. It is therefore obvious that paths with extreme (minimal) proper time
\begin{equation}\label{minimalerWeg}
\int_0^1\frac{\mathrm d\tau(\lambda)}{\mathrm d\lambda}\,\mathrm d\lambda
=\int_{\tau_1}^{\tau_2}\mathrm d\tau
\end{equation}
are proportional to a suitable action function for free particles. The proper time decreases as the speed of the particles increases. The proportionality factor between extreme time and extreme action has the dimension of energy, obviously the rest energy of the free particle
\begin{equation}\label{freieWirkung}
S:=-m_0c_0^2\int_{\tau_1}^{\tau_2}\mathrm d\tau\ist{minimalerWeg}
-m_0c_0^2\int_{t_1}^{t_2}\frac{\mathrm d\tau(t)}{\mathrm dt}\,\mathrm dt
\end{equation}
{The} sign was chosen negative so that the Lagrange function
\begin{equation}\label{freieLagrange}
L:=-m_0c_0^2\frac{\mathrm d\tau(t)}{\mathrm dt}
\ist{betagammatau}-\frac{m_0c_0^2}{\gamma}
\ist{betagammatau}-m_0c_0^2\sqrt{1-\beta^2}
=-m_0\gamma c_0^2(1-\beta^2)
\end{equation}
increases with increasing momentum of the particle. The components of the canonically conjugated momentum follow in the Lagrange description to
\begin{equation}\label{freieImpulse}
p_i:=\frac{\partial L}{\partial v_i}\ist{freieLagrange}m_0\gamma\,v_i.
\end{equation}
{The} Hamiltonian results in
\begin{equation}\label{freieHamilton}
H:=\vec p\,\vec v-L\iist{freieImpulse}{freieLagrange}m_0\gamma c_0^2.
\end{equation}
{For} point-like electrons with the world line $\vec x_e(t)$, the Lagrange density $\mathcal L$, energy density $\mathcal E$ and momentum density $\vec\pi$ are
\begin{equation}\begin{aligned}\label{freieDichten}
&\mathcal L(x)\ist{freieLagrange}
-\frac{m_0c_0^2}{\gamma}\delta^3(\vec x-\vec x_e(t)),\\
&\mathcal E(x)\ist{freieHamilton}m_0\gamma\,c_0^2\delta^3(\vec x-\vec x_e(t)),\\
&\vec\pi(x)\ist{freieImpulse}m_0\gamma\,\vec v\,\delta^3(\vec x-\vec x_e(t))
\end{aligned}\end{equation}

The energy--momentum tensor, which contains the energy density $\mathcal E:=\Theta^{00}(x)$ and momentum density $\vec\pi:=\frac{1}{c_0}\Theta^{0i}(x)$, is
\begin{equation}\label{freieSpannung}
\Theta^{\mu\nu}(x)\iist{freieDichten}{UVonbeta}
\frac{m_0}{\gamma}\,u^\mu(x)u^\nu(x)\,\delta^3(\vec x-\vec x_e(t)).
\end{equation}
{The} momentum of the particle consequently results as a spatial integral over the space--time components of $\Theta^{\mu0}(x)=\Theta^{0\mu}(x)$
\begin{equation}\label{ImpulsAbgeleitet}
p^\mu\iist{freieSpannung}{freieDichten}\frac{1}{c_0}\int_\Sigma\Theta^{\mu0}(x)\,
\mathrm d^3\sigma.
\end{equation}
{As} can be seen from the transition from the integrated quantities in Equations~(\ref{freieLagrange})--(\ref{freieHamilton}) to the densities (\ref{freieDichten}), the integration takes place over the three-dimensional space $\Sigma$ in which the velocities $\vec v$ are determined, i.e., in principle, over any three-dimensional space-like volume $\Sigma$. Precisely this arbitrariness is obviously one of the characteristics of a~particle.

\section{Electrons in Maxwell's Field Model}\label{SecElektronMaxw}
We calculate energy and momentum for an extended classical electron of charge $e=-e_0$ and describe it from different reference systems, in a reference system $\stackrel{\circ}{\Sigma}$ in which the electron is at rest and in a reference system $\Sigma$ in which the electron moves with a velocity $\vec v=c\vec\beta$, see Figure~\ref{mitbewegt}.

Like Abraham~\cite{Abraham1903}, we start from the idea that the mass of the
electron is purely electromagnetic in nature. The electron at rest is described solely by an electric field $\vec E$, which is measured in V/m, and a moving electron additionally by a magnetic field $\vec B$ in T~=~Vs/m$^2$. In relativistic notation, electric and magnetic fields can be combined to form the field strength tensor in SI notation
\begin{equation}\label{FaradayTensor}
F^{\mu\nu}=\begin{pmatrix}
0&-\frac{E_x}{c_0}&-\frac{E_y}{c_0}&-\frac{E_z}{c_0}\\
\frac{E_x}{c_0}&0&-B_z&B_y\\
\frac{E_y}{c_0}&B_z&0&-B_x\\
\frac{E_z}{c_0}&-B_y&B_x&0\end{pmatrix}
\end{equation}
where $c_0$ is the speed of light in the vacuum, which results from the dielectric constant $\varepsilon_0$ and the permeability constant $\mu_0$ of the vacuum by $c_0^2\,\varepsilon_0\,\mu_0=1$. We therefore use the usual Lagrangian density of electrodynamics
\begin{equation}\label{LagrEdyn}
\mathcal L:=-\frac{1}{4\mu_0}F_{\mu\nu}F^{\mu\nu}\ist{FaradayTensor}
\frac{1}{2}\Big(\varepsilon_0\vec E^2-\frac{1}{\mu_0}\vec B^2\Big)
\end{equation}
and the metric tensor $\eta_{\mu\nu}=\mathrm{diag}(1,-1,-1,-1)$. Due to the translation symmetry and after adding a four-divergence term, this leads to the symmetric energy--momentum {tensor} 
\begin{equation}\label{symmEITEdyn}
\Theta^{\mu\nu}=-\frac{1}{\mu_0}\eta^{\mu\kappa}F_{\kappa\lambda}F^{\nu\lambda}
+\frac{1}{4\mu_0}\eta^{\mu\nu}F_{\kappa\lambda}F^{\kappa\lambda},
\end{equation}
see Equation (8.185) of Ref.~\cite{Chaichian2016}. Its elements are in detail
\begin{equation}\label{EneDicht}
\Theta^{00}=:\mathcal E\ist{symmEITEdyn}
\frac{\varepsilon_0}{2}(\vec E^2+c_0^2\vec B^2),
\end{equation}
\begin{equation}\label{ImpDicht}
\Theta^{0i}=\Theta^{i0}\ist{symmEITEdyn}c_0\varepsilon_0\,\vec E\times\vec B,
\end{equation}
\begin{equation}\label{SpannungsDicht}
\Theta^{ij}\ist{symmEITEdyn}-\varepsilon_0\,(E_iE_j+c_0^2B_iB_j)
+\delta_{ij}\frac{\varepsilon_0}{2}(\vec E^2+c_0^2\vec B^2).
\end{equation}

The four-momentum of a field distribution generally depends on the reference system $\Sigma$ in which the field strengths are determined. A reference system can be defined in a Lorentz covariant manner as the three-dimensional Euclidean space that is orthogonal to a velocity vector $u^\mu$ and is usually written as
\begin{equation}\label{3DRaum}
\mathrm d\sigma^\mu:=\frac{u^\mu}{c_0}\,\mathrm d^3\sigma
\end{equation}
{A} Lorentz covariant definition of the four-momentum is obtained by
\begin{equation}\label{DefViererimp}
P^\mu:=\frac{1}{c_0}\int_\Sigma\Theta^{\mu\nu}\,\mathrm d\sigma_\nu.
\end{equation}

In his 1903 paper, Abraham describes electrons on a purely electromagnetic basis with a homogeneous spherically symmetrical charge distribution $\rho(r)$ for electrons at rest. He poses the question: Can the inertia of the electron be completely described by the dynamic effect of its electromagnetic field? It turns out through the 4/3 problem that this question must ultimately be answered in the negative. We will draw conclusions from this failure.

\subsection{Self-Energy of the Classical Electron}
The electric field of a point charge $e$ at rest at the origin results in the SI according to Gauss's law to
\begin{equation}\label{EPunkt}
\vec E_\infty=\frac{e}{4\pi\varepsilon_0}\frac{\vec e_r}{r^2}.
\end{equation}
{As} Equation~(\ref{Eruhend}) will show for the limiting case $r_0\to0$, such a field strength is unrealistic, since it leads to an infinite self-energy $E_e(0)$ of the charge $e$ at rest, i.e., an energy that is infinitely greater than the electron mass requires. The more realistic assumption that the electron charge is distributed on a homogeneously charged sphere of radius $r_0$ or the surface charge of a conducting sphere, as used by Abraham in 1902 on page 147 of Ref.~\cite{Abraham1903}, leads to a finite self-energy. In the following, we prefer a regularized form of the electric field strength $\vec E(0)$ for a resting classical electron, which does not result in a kink or jump in the density of an extended charge distribution,
\begin{equation}\label{Ereg}
\vec E(0):=\frac{e}{4\pi\varepsilon_0}\frac{\vec e_r}{r^2+r_0^2},
\end{equation}
as proposed by Schwinger in Ref.~\cite{Schwinger1983}. In Equation~(\ref{Ereg}) it may be irritating that the field strength at the origin has no defined direction. However, this only shows that the electric field strength is defined as the electric flux density on space--time surfaces, which can have different directions starting from the origin. As required, expression~(\ref{Ereg}) results in a finite energy density everywhere in the system at rest
\begin{equation}\label{enedicht}
\mathcal E_0(\vec r)\ist{EneDicht}
\frac{\varepsilon_0}{2}\vec E^2(0)\ist{Ereg}
\frac{\alpha_f\hbar c_0}{8\pi r_0^4}\frac{1}{(1+\rho^2)^2}\quad\textrm{with}
\quad\alpha_f:=\frac{e_0^2}{4\pi\varepsilon_0\hbar c_0},\quad\rho:=\frac{r}{r_0}.
\end{equation}
with a total energy, the so-called self-energy, of the charge $e$ that results from integration over the three-dimenional space,
\begin{equation}\label{Eruhend}
E_e(0)\ist{DefViererimp}4\pi \int_0^\infty r^2\mathcal E_0(\vec r)\mathrm dr\ist{enedicht}
\frac{\alpha_f\hbar c_0}{r_0}\int_0^\infty\frac{\rho^2}{2(1+\rho^2)^2}\mathrm d\rho
=\frac{\alpha_f\hbar c_0}{r_0}\frac{\pi}{8}=:m_sc_0^2,
\end{equation}
and gives the self-energy $m_sc_0^2$ of the classical electron as a function of $r_0$. Adjusting this self-energy to the physical value leads to $r_0=1.1066$\,fm.

From Equation (\ref{Eruhend}), the known instability of the classical electron can be seen. Its mass decreases with $1/r_0$, it dissolves, its radius parameter $r_0$ increases indefinitely. The reason for this expansion of the core region is easy to detect. Because of the four Lorentz indices in $F_{\mu\nu}F^{\mu\nu}$, the energy density is proportional to $r_0^{-4}$, the spatial integral only grows with $r_0^3$, so overall the $r_0^{-1}$ behavior of Equation~(\ref{Eruhend}) results.

\subsection{Energy of the Moving Electron}
To describe an electron moving in $\Sigma$, we start from the transformation of
the coordinates $\stackrel{\circ}{x}:=(c_0\!\!\stackrel{\circ}{t},\stackrel{\circ}{\vec r)}$ in the comoving frame $\stackrel{\circ}{\Sigma}$ and transform to $\Sigma$ in which the electron moves with $\vec\beta$
\begin{eqnarray}\begin{aligned}\label{txbewegt}
c_0\!\stackrel{\circ}{t} &=\gamma c_0t-\gamma\vec\beta\vec r,\\
\stackrel{\circ}{\vec r}&=\gamma\vec r_\parallel+\vec r_\perp
-\gamma\vec\beta c_0t,
\end{aligned}\end{eqnarray}
{The} Lorentz transformation of the field strength tensor $F^{\mu\nu}$ is
\begin{eqnarray}\begin{aligned}\label{EBbewegt}
&\vec E(\vec\beta)=\vec E_\parallel(0)+\gamma\vec E_\perp(0)\quad
\textrm{with}\quad\vec E_\parallel:=\frac{\vec\beta(\vec\beta\vec E)}{\beta^2},
\quad\vec E_\perp:=\vec E-\vec E_\parallel,\\
&c_0\vec B(\vec\beta)=\gamma\vec\beta\times\vec E(0).
\end{aligned}\end{eqnarray}
{We} can interpret the evaluation of the energy in Equation~(\ref{Eruhend}) in such a way that the energy of the electron at rest regularized according to Equation~(\ref{Ereg}) takes place in the 3D space $\stackrel{\circ}{\Sigma}$, which is orthogonal to the velocity vector $u=\gamma(c,\vec\beta)$, see Figure~\ref{mitbewegt}. We note for further calculations that in $\stackrel{\circ}{\Sigma}$, each of the three electric field components contributes to the energy density~(\ref{EneDicht}) with one-third of $\mathcal E_0(\vec r)$, which is due to the spherical symmetry of the field. Now, however, we consider the energy densities $\mathcal E_{\vec\beta}(\vec r)$ in the 3D space $\Sigma$, which is orthogonal to $(1,\vec 0)$ and contains other space--time points than $\stackrel{\circ}{\Sigma}$, see Figure~\ref{mitbewegt}. We list the contributions of the field components $\vec E_\parallel ,\vec E_\perp, \vec B_\parallel, \vec B_\perp$ according to Equation (\ref{EBbewegt}) in order \footnote{We point out that this calculation, which was carried out in analogy to Abraham~\cite{Abraham1903}, is an exact calculation according to  definition~(\ref{DefViererimp}) of the four-momentum and not, as Rohrlich~\cite{rohrlich2007} writes before his Equation~(16): ``We can summarize this discussion by saying that  definition (5) is incorrect''. With  definition (5), Rohrlich refers to the Abraham-Lorentz definition of the energy of a moving electron, which corresponds to Equation~(\ref{eneBewSigma})}.
\begin{equation}\label{eneBewSigma}
E_e(\beta)\ist{DefViererimp}\int_\Sigma\mathrm d^3\sigma\Theta^{00}(\beta)
\ist{EneDicht}\int_\Sigma\mathrm d^3\sigma\mathcal E_{\vec\beta}(\vec r)
\iist{EneDicht}{EBbewegt}\int_{\Sigma}\mathrm d^3\sigma\frac
{\mathcal E_0(\stackrel{\circ}{\vec r})}{3}(1+2\gamma^2+0+2\gamma^2\beta^2).
\end{equation}
{In} this calculation, we have used that $\vec E_\parallel$ contributes unchanged with one third of the energy density of the electron at rest. The contributions of the two orthogonal electric field components are given a factor $\gamma^2$. $\vec B_\parallel$ and consequently its contribution vanishes. The two orthogonal magnetic field components are proportional $\beta^2\gamma^2$ according to Equation~(\ref{EBbewegt}). From an expansion up to the order $\beta^2$, i.e., $\gamma^2\approx1+\beta^2$, Abraham read in Equation~(15e) of Ref.~\cite{Abraham1903} that the ``kinetic'' magnetic energy contributions proportional to $\beta^2$, $W_\mathrm{m}\propto2\gamma^2\beta^2\approx2\beta^2$, is related to the ``static'' electrical energy contributions of a Lorentz-contracted electron $W_\mathrm{e}\propto1+2\gamma^2\approx3+2\beta^2$ by
\begin{equation}\label{AbrahamBez}
W_\mathrm{m}\ist{eneBewSigma}\approx\frac{2\beta^2}{3+2\beta^2}\,W_\mathrm{e}
\approx\frac{2\beta^2}{3}\,W_\mathrm{e}
\approx\frac{4}{3}\frac{\beta^2}{2}\,W_\mathrm{e},
\end{equation}
and thus the factor $4/3$ appeared for the first time.

Due to the time independence of the electric field strength in the comoving system, it was possible in Equation~(\ref{eneBewSigma}) to read off the field values in the comoving reference system $\stackrel{\circ}{\Sigma}$ instead of in $\Sigma$, see Figure~\ref{mitbewegt}. Since we have already integrated in Equation~(\ref{Eruhend}) over the energy density in the comoving system $\stackrel{\circ}{\Sigma}$, it makes sense to carry out the integration in Equation~(\ref{eneBewSigma}) via $\stackrel{\circ}{\Sigma}$, whereby we take into account the Lorentz contraction of the moving electron according to Equation~(\ref{txbewegt}). The total energy of the moving classical electron thus results in
\begin{equation}\label{eneBew0}
E_e(\beta)\iist{eneBewSigma}{txbewegt}\frac{1}{\gamma}\int_{\stackrel{\circ}{\Sigma}}
\mathrm d^3\stackrel{\circ}{\sigma}\frac{\mathcal E_0(\stackrel{\circ}{\vec r)}}
{3}(1+2\gamma^2+0+2\underbrace{\gamma^2\beta^2}_{\gamma^2-1})
\ist{Eruhend}\frac{E_e(0)}{3\gamma}(4\gamma^2-1),
\end{equation}
which does not have the form~(\ref{pvonbeta}) expected for particles. For $\gamma=1$ the expression is correct, but for $\gamma\to\infty$ the energy increases to $4/3$ of the expected value.

\subsection{Momentum of the Moving Electron}
According to the particle interpretation, the momentum of the uniformly moving classical electron results from the $\Sigma$ integration via the stress tensor or the Poynting vector
\begin{equation}\begin{aligned}\label{ImpBew}
\vec P_e(\vec\beta)&:=\frac{1}{c_0}\int_\Sigma\mathrm d^3\sigma\,\Theta^{0i}
(\vec\beta)\ist{ImpDicht}\varepsilon_0\int_\Sigma\mathrm d^3\sigma\vec E
(\vec\beta)\times\vec B(\vec\beta)=\\
&\ist{EBbewegt}\frac{\varepsilon_0}{c_0}\gamma
\int_\Sigma\mathrm d^3\sigma\left[\vec E_\parallel(0)+\gamma\vec E_\perp(0)\right]
\times\left[\vec\beta\times\vec E_\perp(0)\right].
\end{aligned}\end{equation}
In the transformation to $\stackrel{\circ}{\Sigma}$, we again take into account the Lorentz contraction
\begin{equation}\begin{aligned}\label{ImpBew0}
\vec P_e(\vec\beta)&\iist{ImpBew}{txbewegt}
\frac{\varepsilon_0}{c_0}\int_{\stackrel{\circ}{\Sigma}}\mathrm d^3\stackrel{\circ}
{\sigma}\left[\vec E_\parallel(0)+\gamma\vec E_\perp(0)\right]\times
\left[\vec\beta\times\vec E_\perp(0)\right]=\\&=\frac{\varepsilon_0}{c_0}
\int_{\stackrel{\circ}{\Sigma}}\mathrm d^3\stackrel{\circ}{\sigma}
\left[-\vec E_\perp(0)\left(\vec E_\parallel(0)\vec\beta\right)
+\gamma\vec\beta\vec E_\perp^2(0)\right].
\end{aligned}\end{equation}
The first summand in the integrand of the last expression shows momentum densities normal to the velocity and thus internal stresses in the classical electron. At points that are mirror-symmetrical to the velocity vector $\vec\beta$, $E_\perp(0)$ points in the opposite directions, so their contributions cancel each other out and do not contribute to the total momentum $\vec P_e(\vec\beta)$. Since the two orthogonal field components $\vec E_\perp(0)$ in the resting electron each contribute one-third of the field energy, the second contribution provides the 4/3 factor already known from Equation~(\ref{AbrahamBez})
\begin{equation}\label{ImpBew1}
\vec P_e(\vec\beta)\ist{ImpBew0}\vec\beta\gamma\frac{4}{3}
\frac{\varepsilon_0}{c_0}\int_{\stackrel{\circ}{\Sigma}}
\mathrm d^3\stackrel{\circ}{\sigma}\frac{\vec E^2(0)}{2}\ist{Eruhend}\vec\beta
\gamma\frac{4}{3}\frac{E_e(0)}{c_0}\ist{Eruhend}\vec v\gamma\frac{4}{3}m_s.
\end{equation}
{Both} calculations, (\ref{AbrahamBez}) and (\ref{ImpBew1}), thus show that the mass of the electron contributing to the momentum and the kinetic energy is greater by a factor of 4/3 than results from the electrical field energy $E_e(0)$ for the electron at rest in Equation~(\ref{eneBew0}). This would mean a discrepancy between inertial and gravitational mass. To eliminate this contradiction, Poincaré introduced a negative pressure~\cite{poincare1905}, which has to balance the exploding tendency of the electron.

\subsection{Lorentz Transformed Four-Momentum of the Electron at Rest}
The situation is different for the four-momentum of the electron at rest in $\stackrel{\circ}{\Sigma}$ when its four-momentum $P^\mu(\vec\beta)$ is expressed in coordinates of $\Sigma$. As Rohrlich~\cite{rohrlich1960} has clearly shown, the correct expression results due to the consistency of special relativity
\begin{equation}\label{ViererSigma0}
P^\mu(\vec\beta)\iist{DefViererimp}{3DRaum}\frac{1}{c_0}\int_{\stackrel{\circ}{\Sigma}}\mathrm d^3\stackrel{\circ}{\sigma}\Theta^{\mu\nu}_\Sigma\beta_\nu
\iist{eneRuhSigma0}{ImpRuhSigma0b}\frac{E_e(0)}{c_0}(\gamma,\vec\beta\gamma),
\end{equation}
whereby it should be noted that the field values calculated in the laboratory system $\Sigma$ are integrated over the comoving world volume $\stackrel{\circ}{\Sigma}$.

In detail, this results in
\begin{equation}\begin{aligned}\label{eneRuhSigma0}
P^0(\vec\beta)&\ist{ViererSigma0}\frac{\gamma}{c_0}\int_{\stackrel{\circ}{\Sigma}}
\mathrm d^3\stackrel{\circ}{\sigma}(\Theta^{00}_\Sigma-\beta_i\Theta^{0i}_\Sigma)
\ist{txbewegt}\frac{\gamma^2}{c_0}\int_\Sigma
\mathrm d^3\sigma(\Theta^{00}_\Sigma-\beta_i\Theta^{0i}_\Sigma)=\\
&\iist{eneBewSigma}{ImpBew}\gamma^2\frac{E_e(\beta)}{c_0}-\gamma^2\vec\beta
\vec P_e(\vec\beta)\iist{eneBew0}{ImpBew1}\gamma(4\gamma^2-1)\frac{E_e(0)}{3c_0}
-\beta^2\gamma^3\frac{4E_e(0)}{3c_0}=\gamma E_e(0)
\end{aligned}\end{equation}
\begin{equation}\begin{aligned}\label{ImpRuhSigma0}
P^i(\vec\beta)&\ist{ViererSigma0}\frac{\gamma}{c_0}\int_{\stackrel{\circ}{\Sigma}}
\mathrm d^3\stackrel{\circ}{\sigma}(\Theta^{i0}_\Sigma-\beta_j\Theta^{ij}_\Sigma)=\\
&\iist{SpannungsDicht}{ImpBew}\gamma^2P_e^i(\vec\beta)-\frac{\gamma}{c_0}\beta_j
\int_{\stackrel{\circ}{\Sigma}}\mathrm d^3\stackrel{\circ}{\sigma}
\{-\varepsilon_0\,[E_i(0)E_j(0)+c_0^2B_i(0)B_j(0)]+\\&\hspace{20mm}
+\delta_{ij}\frac{\varepsilon_0}{2}[\vec E^2(0)+c_0^2\vec B^2(0)]\}\\
&\ist{EBbewegt}\gamma^2P_e^i(\vec\beta)+\frac{\gamma}{c_0}\beta\varepsilon_0\,
\int_{\stackrel{\circ}{\Sigma}}\mathrm d^3\stackrel{\circ}{\sigma}
[E_i(0)E_\parallel(0)+c_0^2B_i(0)\underbrace{B_\parallel(0)}_{0}]-\\&\hspace{20mm}
-\frac{\gamma}{c_0}\beta_i\frac{\varepsilon_0}{2}\int_{\stackrel{\circ}{\Sigma}}
\mathrm d^3\stackrel{\circ}{\sigma}[\vec E^2(0)+c_0^2\vec B^2(0)]
\end{aligned}\end{equation}
i.e.,
\begin{equation}\begin{aligned}\label{ImpRuhSigma0b}
P_\parallel(\vec\beta)&\ist{ImpRuhSigma0}\gamma^2P_e^\parallel(\vec\beta)
+\frac{\gamma}{c_0}\beta\frac{\varepsilon_0}{2}\,
\int_{\stackrel{\circ}{\Sigma}}\mathrm d^3\stackrel{\circ}{\sigma}[E_\parallel^2(0)
-2E_\perp^2(0)-2c_0^2\vec B^2_\perp(0)]=\\
&\iist{ImpBew1}{EBbewegt}\beta\gamma^3\frac{4}{3}\frac{E_e(0)}{c_0}
+\frac{\gamma}{c_0}\beta\,\frac{E_e(0)}{3}
[1-2\gamma^2-2\beta^2\gamma^2]=\beta\gamma\frac{E_e(0)}{c_0}.
\end{aligned}\end{equation}
and
\begin{equation}\begin{aligned}\label{ImpRuhSigma0c}
P_\perp(\vec\beta)&\ist{ImpRuhSigma0}\frac{\gamma}{c_0}\beta\varepsilon_0\,\int
_{\stackrel{\circ}{\Sigma}}\mathrm d^3\stackrel{\circ}{\sigma}E_\parallel(0)E_\perp(0)=0.
\end{aligned}\end{equation}
{The} result~(\ref{ViererSigma0}) has nothing to do with the 4/3 problem. It only shows that the four-vector $(1,\vec0)\,E_e(0)/c_0$ in $\stackrel{\circ}{\Sigma}$ can be transformed by a Lorentz transformation to $(\gamma,\vec\beta\gamma)\,E_e(0)/c_0$ in $\Sigma$. It is important that the space-like volume that is integrated is also correctly transformed.

\section*{Acknowledgement}
I thank Janos Polonyi for pointing out the importance of solving the problems of the classical electron, and Christoph Adam and Martin Suda for suggesting corrections to the first version of the manuscript.

\printbibliography[title={References}]
\end{document}